\documentclass[10pt,aps,prb,twocolumn,tightenlines,letterpaper,superscriptaddress,nofootinbib,floatfix]{revtex4-2}
\usepackage{amssymb,latexsym}
\usepackage{amsmath,amsbsy,bbm}
\usepackage{epsfig,bm,color,bbold}
\usepackage{graphicx}
\usepackage{xcolor}
\usepackage{cancel}
\usepackage{enumitem}

\usepackage{lineno}

\begin{document}

\def\a{{\alpha}}
\def\be{{\beta}}
\def\d{{\delta}}
\def\D{{\Delta}}
\def\P{{\Pi}}
\def\p{{\pi}}
\def\e{{\varepsilon}}
\def\ep{{\epsilon}}
\def\g{{\gamma}}
\def\k{{\kappa}}
\def\l{{\lambda}}
\def\L{{\Lambda}}
\def\m{{\mu}}
\def\n{{\nu}}
\def\o{{\omega}}
\def\O{{\Omega}}
\def\S{{\Sigma}}
\def\s{{\sigma}}
\def\t{{\tau}}
\def\x{{\xi}}
\def\X{{\Xi}}
\def\z{{\zeta}}

\def\ol#1{{\overline{#1}}}
\def\c#1{{\mathcal{#1}}}
\def\b#1{{\bm{#1}}}
\def\eqref#1{{(\ref{#1})}}

\def\wt#1{{\widetilde{#1}}}

\def\ed#1{{\textcolor{magenta}{#1}}}
\def\edd#1{{\textcolor{cyan}{#1}}}


\author{Prabal \surname{Adhikari}}
\email[]{$\texttt{prabal.adhikari.physics@proton.me}$}
 \affiliation{
Physics Department,
        Faculty of Natural Sciences and Mathematics,
        St.~Olaf College,
        Northfield, MN 55057, USA}
\affiliation{Kavli Institute for Theoretical Physics, 
	University of California, 
	Santa Barbara, CA 93106, USA.}

\author{Sona~Baghiyan}
\email[]{$\texttt{baghiyan@wisc.edu}$}
 \affiliation{
        St.~Olaf College,
        Northfield, MN 55057, USA}
\affiliation{Department of Physics,
        University of Wisconsin,
        Madison, WI 53706, USA}
       

\author{Rayn~Samson}
\email[]{$\texttt{samson2@stolaf.edu}$}
 \affiliation{
Physics Department,
        Faculty of Natural Sciences and Mathematics,
        St.~Olaf College,
        Northfield, MN 55057, USA}

\title{Finite Volume Thermodynamics of an Ideal Gas in a Periodic Box }

\date{\today}
\begin{abstract}
Approach to the thermodynamic limit of a non-relativistic ideal gas in a periodic box is investigated. 
The single particle wave function obeys twisted boundary condition, $\psi(L)=e^{i\theta}\psi(0)$ for which the free particle spectrum is constructed in terms of the twist angle, $\theta$. 
The exact density of states is utilized to construct finite-size corrections of thermodynamic observables.  Leading finite volume corrections in the free energy do not arise due to the boundary  -- its implication for mixing entropy is examined. Finite volume corrections to the average energy, its fluctuations and the pressure are also examined with corrections arising exclusively through the boundary condition. However, the equation of state, the ratio of pressure to energy density, remains unmodified by the boundary.
\end{abstract}

\maketitle
\section{Introduction}
\label{sec:introduction}
An ideal gas is a quintessential example in introductory physics, thermodynamics and statistical mechanics courses. 
It is commonly introduced through a classical treatment based on kinetic theory with a focus on the derivation of the ideal gas law and a preview of the equipartition theorem~\cite{mello2010}.

The discussion is typically upgraded in statistical physics courses through the utilization of the Boltzmann distribution (from which the equipartition theorem is constructed) and the canonical ensemble for which the partition function is the fundamental object. 
At inverse temperature, $
\be$, see Eq.~\eqref{eq:de-Broglie-wavelength}
the single particle partition function of an ideal gas in a one-dimensional box is an exponential sum
\begin{align}
\label{eq:Ze}
Z_{1}=\sum_{n}\,\O(\e_{n})\,e^{-\be \e_{n}}.
\end{align}
The quantity $\O(\e_{n})$ in the sum accounts for the degeneracy associated with the single particle eigenstate $|n\rangle$ with energy $\e_{n}$. The spectrum itself is determined through the Schr\"{o}dinger equation using a suitable set of boundary conditions. 
The most common choices are Dirichlet boundary condition for which the wave function vanishes on the edges of the box and periodic boundary condition for which the wave function is continuous and differentiable on the edges.  

In the standard canonical ensemble treatment of the ideal gas, the sum in the single particle partition function is replaced by an integral\footnote{While the sum bounds  and integral bounds (both $n$ and $k$) have been omitted, they depend on the choice of boundary condition. For Dirichlet boundary condition, the lower and upper bounds on the sum are $1$ and $\infty$ while for periodic boundary condition, the bounds are $\mp\infty$. The integrals possess identical bounds to the sum for periodic boundary condition. However, for Dirichlet boundary condition, the integral bounds are $0$ and $\infty$.}~\cite{stutz1968,markworth1971,fernandez2015}
\begin{align}
\label{eq:density-of-states}
\sum_{n}\rightarrow \smallint dn=\smallint dk\, g(k)
\end{align}
in order to access the thermodynamic limit, i.e. the limit in which the number of gas particles, $N$, approaches infinity at fixed gas density\footnote{For instance, see Refs.~\cite[Chapter 6]{schroeder2021introduction} and \cite[Chapter 21]{blundell2010concepts}.}. The quantity $g(k)$ in the integrand is the $k$-space density of states with $g(k)dk$ characterizing the number of states within a window of wave numbers between $k$ to $k+dk$.\footnote{It is also encountered by undergraduate students in a modern physics course~\citep[Chapter 3]{townsend2000modern}.} The replacement of the sum by an integral leads to the inevitable student question, \\

\begin{figure}[t!]
\centering
\includegraphics[width=0.45\textwidth]{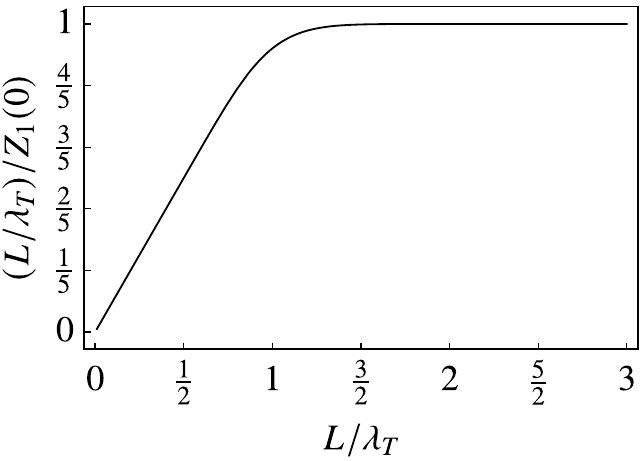}\hspace{40pt}
\caption{
Plot of the ratio of the approximate single particle partition function, $L/\l_{T}$, to the exact single-particle partition function, $Z_{1}(0)$, in one spatial dimension, see Eq.~\eqref{eq:single-particle-partition-function}, as a function of box size (normalized by the de Broglie wavelength), $L/\l_{T}$.}
\label{fig:Z}
\end{figure}

\indent\textit{``How good is the approximation?"}.\\
\\
\noindent
The inquiry can be addressed through either a numerical justification or a heuristic response. A numerical justification compares the ratio of the approximated partition function with its exact counterpart. An illustration presented in Fig.~\ref{fig:Z} exhibits excellent convergence of the single particle partition function in a one dimensional periodic box of approximate size $L\gtrsim 2\l_{T}$. $\l_{T}$ is the de Broglie wavelength
\begin{align}
\label{eq:de-Broglie-wavelength}
\l_{T}&=\hbar\sqrt{\frac{2\p \be}{m}}&
\be&=\frac{1}{k_{B}T}\,.
\end{align}
In the heuristic approach, one notes that the energy levels, $\e_{n}$, are proportional to Planck's constant and therefore closely spaced\footnote{For instance see Ref.~\citep[Chapter 3]{townsend2000modern}.} with the replacement of a sum by an integral producing an excellent approximation for large box sizes and temperature.


While these responses are well-justified, the analysis of the nature of the convergence of thermodynamic observables can be made precise using tools readily available to undergraduate students. In fact, Fourier series and Gaussian integrals suffice in the construction of finite volume corrections to thermodynamic observables. We will conduct such analysis for a periodic boy with twisted boundary condition. $\theta=0$ corresponds to the commonly utilized periodic boundary condition.

The paper is organized as follows: In Section~\ref{sec:boundary-condition}, we discuss the origin of twisted boundary condition in a periodic box and construct the single particle spectrum as a function of the twist angle $\theta$. In Section~\ref{sec:ideal-gas-canonical-ensemble}, we utilize the exact density of states to reformulate the single particle partition function that makes it amenable to a finite volume analysis. In Section~\ref{sec:finite-volume-thermodynamics}, we construct thermodynamic observables of an ideal gas with a particular focus on the approach to the thermodynamic limit. In this context, we study the free energy per particle  and mixing entropy followed by an investigation of the average energy, its fluctuations, the pressure and the equation of state. We conclude with exercises designed to aid the mastery of the techniques discussed in the paper. In an accompanying Appendix~\ref{app:fourier-series}, we discuss the Fourier series of a periodic $\delta$-function.


\section{Boundary Condition \& Single Particle Spectrum}
\label{sec:boundary-condition}
Consider a one-dimensional periodic box, $x\in [\,0,L\,)$, with twisted boundary condition imposed on the time-independent wave function, $\psi(x)$,
\begin{align}
\label{eq:twisted-boundary-condition}
\psi(L)=e^{i\theta}\psi(0)\, .
\end{align}
While the wave function itself is only periodic upto the twist angle, $\theta$, the probability density, $|\psi(x)|^{2}$,  is periodic. 

The twist angle arises due to the requirement that the momentum operator is self-adjoint\footnote{The standard requirement for Hermiticity discussed in introductory quantum mechanics courses is not sufficient to ensure that the domain of the momentum operator, $\hat{p}$, and its hermitian conjugate, $\hat{p}^{\dagger}$ are identical. For a thorough and accessible discussion of these issues, see Ref.~\cite{bonneau2001self}.}
\begin{align}
\langle \phi | \hat{p} | \psi\rangle&=\langle \phi | \hat{p}\, \psi\rangle=\langle \hat{p}\,\phi | \psi\rangle\, .
\end{align}
Using the differential form of the momentum operator in position space, $-i\hbar\frac{\partial}{\partial x}$, the difference between the latter two matrix elements defined as
\begin{align}
\D(\hat{p})&\equiv\langle \phi | \hat{p}\, \psi\rangle-\langle \hat{p}\,\phi | \psi\rangle
\end{align}
is generally non-zero due to boundary contributions as is readily determined through an integration-by-parts
\begin{align}
\D(\hat{p})=&-i\hbar\,\Big.\phi^{*}(x)\psi(x)\Big|_{0}^{L}\,.
\end{align}
The difference must vanish for the momentum operator to be self-adjoint. Requiring that the wave function satisfies the condition
\begin{align}
\psi(L)&=\l\, \psi(0)
\end{align}
where $\l$ is a complex valued parameter, leads to the requirement, $\l \l^{*}=1$.
The constraint is equivalent to the twisted boundary condition of Eq.~\eqref {eq:twisted-boundary-condition} upon writing $\l$ as
\begin{align}
\l &= e^{i\theta}&
\theta\in [\,0,2\p\,)\, .
\end{align}
Note that the commonly utilized periodic boundary condition is a subset of twisted boundary condition with zero twist angle, $\theta=0$. 

\begin{figure}[t!]
\centering
\includegraphics[width=0.48\textwidth]{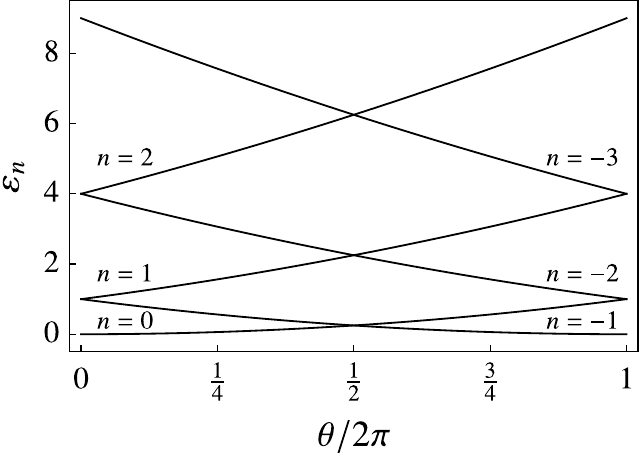}\hspace{40pt}
\caption{
Plot of the spectrum, $\e_{n}(\theta)$, as a function of the twist angle, $\theta$ for $n=0,\pm1,\pm2, 3$.}
\label{fig:spectrum}
\end{figure}

Prior to constructing the momentum eigenspectrum, it is worth briefly considering the free-particle Hamiltonian, $\hat{H}$. A self-adjoint requirement on the operator produces upon performing two integration-by-parts
\begin{align}
\D(\hat{H})&=\frac{\hbar^{2}}{2m}\Big.\big[\partial_{x}\phi^{*}(x)\psi(x)-\phi^{*}(x)\partial_{x}\psi(x)\big]\Big|_{0}^{L}\, .
\end{align}
For twisted boundary condition, $\D(\hat{H})$ indeed vanishes ensuring that $\hat{H}$ is self-adjoint. Additionally, the free particle Hamiltonian commutes with the momentum operator. Imposing twisted boundary condition on momentum eigenstates (in position space) leads to 
\begin{align}
\psi_{n}(x)=\frac{1}{\sqrt{L}}e^{ik_{n}(\theta)x}\,,
\end{align}
where $k_{n}$ is the quantized wave number for all integer $n$
\begin{align}
\label{eq:pn}
k_{n}(\theta)=\frac{2\p}{L}\left(n+\frac{\theta}{2\p}\right)\, .
\end{align}
The corresponding energy eigenvalues, $\e_{n}$, required for the construction of the single particle partition function are, therefore,
\begin{align}
\label{eq:spectrum}
\e_{n}(\theta)=\frac{p_{n}^{2}}{2m}=\frac{2\p^{2}\hbar^{2}}{mL^{2}}\left(n+\frac{\theta}{2\p}\right)^{2} .
\end{align}
Unlike periodic boundary condition, positive and negative values of $n$ in general possess different energy eigenvalues. In Fig.~\ref{fig:spectrum}, we plot the spectrum as a function of the twist angle for $n=0,\pm 1, \pm 2, -3$. For zero or positive values of $n$, the spectrum monotonically increases in the range $\theta\in [0,2\p)$ while for negative values of $n$, the spectrum monotonically decreases. For $\theta=0$ and $\p$, the spectrum exhibits a double degeneracy. For $\theta=0$, the excited states are doubly degenerate - in particular the states $\pm n$ are degenerate. For $\theta=\p$, however, the degeneracies occur at each accessible energy and for $n\ge0$ and $-(n+1)$.

The separation between nearest energy eigenvalues
\begin{align}
\e_{n}-\e_{n-1}&=\frac{4\p^{2}\hbar^{2}}{mL^{2}}\left(n-\frac{1}{2}+{\frac{\theta}{2\p}}\right)\,,
\end{align}
is not only a function of $n$ but also the twist angle. As such, the presence of $\theta$ impacts the partition function and thermodynamic observables in a non-trivial manner. Naturally, in the thermodynamic limit, the physics of an ideal gas is unaffected by the choice of boundary condition. However, the approach to the thermodynamic limit~\cite{ghosh2015,styer2004} is impacted.

\section{Partition Function of an Ideal Gas}
\label{sec:ideal-gas-canonical-ensemble}
The partition function of an ideal gas in the canonical ensemble is constructed from the single particle partition function, $Z$,
\begin{align}
\label{eq:Z}
\c{Z}&=\frac{Z^{N}}{N!}&
Z&=Z_{1}^{d}\,
\end{align}
where $d$ is the number of spatial dimensions. The $N!$ ensures that thermodynamic observables are well-defined in the thermodynamic limit. Additionally, it ensures that the mixing entropy of a gas of identical, indistinguishable particles, is zero in the thermodynamic limit~\cite{gibbs1878equilibrium}. Since the partition function, $\c{Z}$, is not separable, it encodes statistical correlations in the ideal gas. 

For twisted boundary condition, the (one-dimensional) single-particle partition function can be recast as
\begin{align}
\label{eq:single-particle-partition-function}
Z_{1}(\theta)&=\sum_{n=-\infty}^{\infty}e^{-\p \left(\frac{1}{L_{T}}\right)^{2}\left(n+{\frac{\theta}{2\p}}\right)^{2}}\ ,
\end{align}
in terms of the dimensionless length, $L_{T}$,
\begin{align}
L_{T}=\frac{L}{\l_{T}}\,.
\end{align}
The form of the partition function is inconvenient for exploration of the approach to the thermodynamic limit. The system size appears in the denominator of the exponent. In the thermodynamic limit, this exponent may not be small due to the numerator, $\p(n+\frac{\theta}{2\p})^{2}$ being comparable in size. Consequently, the structure of the sum is not amenable to an expansion in the limit of large box sizes. 
\subsection{Density of States}
\label{sec:density-of-states}
In order to restructure the sum, we utilize the density of states, $g(k)$, defined in Eq.~\eqref{eq:density-of-states}. It characterizes the (infinitesimal) number of states $dn$
\begin{align}
\label{eq:dn}
dn=g(k)dk
\end{align}
within the range $k$ to $k+dk$. 
The density of states approach is commonly utilized (in the undergraduate physics curriculum) to approximate the single-particle partition function
\begin{align}
\label{eq:Z1-k}
Z_{1}&=\int_{-\infty}^{\infty} dk\,g(k)\, e^{-\be \e(k)}.
\end{align}
The exact density of states\footnote{For a graphical discussion of the exact density of states, see Ref.~\cite{mulhall2014calculating}.} is a periodic sum over Dirac $\d$-functions (a Dirac comb) that is non-zero only at the allowed momentum modes
\begin{align}
\label{eq:density-of-states-exact}
g(k)&=\sum_{n=-\infty}^{+\infty}\d(k-k_{n}).
\end{align}
It is designed to reproduce the single particle partition function, Eq.~\eqref{eq:single-particle-partition-function}, which determines the bulk thermodynamics of an ideal gas. Furthermore, it is periodic in units of
\begin{align}
k_{n}(\theta)-k_{n-1}(\theta)=\frac{2\p}{L}\,,
\end{align}
which is independent of the twist angle as is readily evident from Eq.~\eqref{eq:pn}. Note that for Dirichlet boundary condition, a subject of example exercises in Appendix~\ref{app:example-sheet}, the wave numbers are restricted to positive values and doubly dense. 

It is important that the exact density of states is contrasted with the more commonly utilized (approximate) density of states, $g_{0}(k)$, which reproduces the partition function that is utilized for the construction of thermodynamic observables. In the standard treatment, 
\begin{align}
g_{0}(k)dk\equiv\frac{1}{\frac{2\p}{L}}dk
\end{align}
counts, albeit only approximately, the number of momentum states within an infinitesimal range $dk$ of wave numbers. For the one dimensional partition function, the density, $g_{0}(k)$, is independent of $k$.  The replacement $g(k)\rightarrow g_{0}(k)$ in Eq.~\eqref{eq:Z1-k} produces the standard result for the single particle partition function, namely $L/\l_{T}$ -- it does not account for finite volume corrections. While this approach is well-motivated heuristically as the quantum energy levels are narrowly spaced, a precise mathematical formulation is indeed possible. This approach has the added advantage that finite volume corrections can be systematically extracted.

\subsection{Reformulation of $Z_{1}(\theta)$}
\label{sec:reformulation-of-Z1}
In order to construct the exact partition function, we first note that the density of states, $g(k)$ is a periodic function. Rewriting $g(k)$ in terms of its dimensionless counterpart leads to
\begin{align}
g(k)&=\frac{L}{2\p}\varrho(\k)]\equiv\frac{L}{2\p}\sum_{n=-\infty}^{\infty}\d(\k-n)\,.
\end{align}
Upon factoring out the $k$-space period, a dimensionless periodic function $\varrho(\k)$ with period one emerges. $\k$ is compact notation for a dimensionless twist-dependent function
\begin{align}
\k(\theta)&=\frac{kL-\theta}{2\p}\,.
\end{align}
Since $\varrho(\kappa)$ is a periodic function, it admits a Fourier decomposition
\begin{align}
\sum_{n=-\infty}^{\infty}\d(\k-n)&=\sum_{\n=-\infty}^{\infty}c_{\n}\,e^{2\p i \k\n }&
c_{\n}&=1\,.
\end{align}
The Fourier coefficients, $c_{\n}$, are determined by the relation
\begin{align}
c_{\n}=\int_{\k_{0}}^{\k_{0}+1}d\k\,\varrho(\k)e^{2\pi i\k\n}\,,
\end{align}
with the choice for $\k_{0}$ arbitrary. The one in the upper limit is the period of $\varrho(\k)$. For period $\l_{\k}$, the coefficient is $\l_{\k}^{-1}$ as is evident upon performing the integral. While $\varrho(\k)$ is a sum over an infinite number of $\d$-functions, only one of them contributes to the integral. For a detailed discussion of the Fourier decomposition of the periodic $\d$-function, consult Appendix~\ref{app:fourier-series}.

Utilizing the reformulated density of states, the single particle partition function takes the form
\begin{align}
Z_{1}(\theta)=\int_{-\infty}^{+\infty}dk\left(\frac{L}{2\p}\sum_{\n=-\infty}^{\infty}e^{i(kL-\theta)\n}\right)e^{-\be\frac{\hbar^{2}}{2m}k^{2}}
\end{align}
where the $\n=0$ contribution to the integral is the approximate density of states, $g_{0}(k)$. The $k$-integral admits a Gaussian structure and is therefore readily performed\footnote{For convenient access, the Gaussian integral utilized in the analysis is
\begin{align}
\int_{-\infty}^{+\infty}e^{-ax^{2}}e^{ibx}&=\sqrt{\frac{\p}{a}}e^{-\frac{b^{2}}{4a}}&
a&>0\,.
\end{align}}
\begin{align}
Z_{1}(\theta)=L_{T}\sum_{\n=-\infty}^{\infty}e^{-i\n\theta}e^{-\p\left(\n L_{T}\right)^{2}}\,.
\end{align}
In this form, the leading contribution to the sum is the dimensionless size, $L_{T}$. The finite volume contributions are organized into higher order contributions by the magnitude of the winding number, $\n$. Consequently, the single particle partition function is best written as a sum over positive values of $\n$,
\begin{align}
\label{eq:Z1-theta-v2}
Z_{1}(\theta)&=L_{T}\left[1+2\sum_{\n=1}^{\infty}\cos(\n\theta)\,e^{-\p\left(\n L_{T}\right)^{2}}\right],
\end{align}
with the resulting structure amenable to the extraction of thermodynamic observables and finite volume corrections.

\section{Finite Volume Thermodynamics}
\label{sec:finite-volume-thermodynamics}
In this section, we utilize the partition function (in the canonical ensemble) to investigate finite size corrections to thermodynamic observables. We conduct an expansion of thermodynamic observables valid in the limit $N\rightarrow \infty$ at fixed gas density, 
\begin{align}
\rho=\frac{N}{L^{d}}\,.
\end{align}
Since the ensemble, by construction, ignores quantum correlations, the gas density must be low compared to the (inverse) deBroglie wavelength, $\rho\l_{T}^{d}\ll 1$.
The condition is achieved by (sufficiently) heating the gas (which reduces the deBroglie wavelength) or alternatively increasing the box size at finite temperature.

%
%
%
%
%

\subsection{Free Energy}
\label{sec:free-energy}
The ideal gas free energy defined in terms of the partition function\footnote{We utilize the convention $\log e=1$.}, $\c{Z}(\theta)$,
\begin{align}
F(\theta)&=-k_{B}T\log \c{Z}(\theta)\,
\end{align}
admits a well-defined thermodynamic limit due to the $N!$ in Eq.~\eqref{eq:Z}.\footnote{For reference, we quote the leading contribution to the free energy
\begin{align}
\lim_{N\rightarrow \infty} \frac{F(\theta)}{N}=k_{B}T\log\left(\frac{\rho \l_{T}^{d}}{e}\right)\,,
\end{align}
a standard result in statistical mechanics.} Boundary corrections naturally depend on the twist angle
\begin{align}
\hspace{-2mm}
\frac{F_{L}(\theta)}{N}&=-dk_{B}T\log\left[1+2\sum_{\n=1}^{\infty}\cos(\n\theta)\,e^{-\p(\n L_{T})^{2}}\right]\,.
\end{align}
The corrections scale exponentially with the system size as is evident upon expanding the logarithm, which, for large system sizes, is a linear function in the sum. For twist angles $\theta\neq\tfrac{\p}{2},\,\tfrac{3\p}{2}$, the finite size correction to the free energy (relative to the thermodynamic limit result) is $\c{O}(e^{-\p L_{T}^{2}})$. For $\theta\neq\tfrac{\p}{2},\,\tfrac{3\p}{2}$, the leading correction arises through winding numbers with magnitude two and therefore $\c{O}(e^{-4\p L_{T}^{2}})$. 

There is a further correction to the free energy, $F_{N}$, that arises due to the $N!$ in the partition function. In order to extract the associated correction, one utilizes the Stirling approximation\footnote{The Stirling expansion, valid for large values of $N$, admits the following structure
\begin{align}
\log N!=N\log \left(\tfrac{N}{e}\right)+\log\sqrt{2\p N}+\c{O}\left(\tfrac{1}{N}\right)
\end{align}}. We quote only the leading correction
\begin{align}
\frac{F_{N}}{N}=k_{B}T\frac{\log \sqrt{2\p N}}{N}+\c{O}\left(\frac{1}{N^{2}}\right)\,.
\end{align}
At fixed density, the correction (per particle) scales as a logarithm times a power law
\begin{align}
\frac{\log N}{N}\sim\frac{\log{L_{T}^{d}}}{L_{T}^{d}}\,.
\end{align}
Evidently, it is the leading finite size correction to the free energy (per particle) in the ideal gas. 

\subsection{Mixing Entropy}
\label{sec:entropy}
In order to appreciate its significance, consider the mixing entropy of two identical gases with $N$ particles each, confined in two halves of a container, each with volume $L^{d}$ and separated by a partition. Each half has entropy $S(N)$ -- upon removing the partition the total entropy changes to a new entropy, $S(2N)$ with the mixing entropy defined as the change in entropy upon removal of the partition 
\begin{align}
S_{\rm mixing}&=S(2N)-2S(N)\,.
\end{align}
The entropy, $S$, defined at fixed system size and particle number,
\begin{align}
S=-\left(\frac{\partial F}{\partial T}\right)_{L,N}\,,
\end{align}
leads to an expression in thermodynamic limit that is extensive.\footnote{The entropy per particle in an ideal gas scales with $N$ in the thermodynamic limit,
\begin{align}
\lim_{N\rightarrow \infty}\frac{S}{N}&=k_{B}\left[-\log\left(\rho\l_{T}^{d}\right)+\left(\frac{d}{2}+1\right)\right]\,.
\end{align}
Consequently the associated mixing entropy is zero.} The mixing entropy in the thermodynamic limit is zero due to extensivity and resolves the Gibbs paradox~\cite{jaynes1992gibbs}. However, accounting for finite size corrections to the entropy per particle
\begin{align}
\frac{S_{N}}{N}&=-k_{B}\frac{\log \sqrt{2\p N}}{N}+\c{O}\left(\frac{1}{N^{2}}\right)\,.
\end{align}
leads to a non-zero mixing entropy. Using the expression, one finds that the mixing entropy scales logarithmically in $N$ or equivalently, logarithmically in system size at fixed density
\begin{align}
S_{2N}-2S_{N}=k_{B}\log\sqrt{\p N}\sim \log L_{T}^{d}\,.
\end{align}
While there is a further twist angle dependent finite size correction
\begin{align}
\frac{S_{L}(\theta)}{N}&=dk_{B}\log\left[1+2\sum_{\n=1}^{\infty}\cos(\n\theta) e^{-\p L_{T}^{2}}\right]\\
&- dk_{B}\frac{2\p\sum_{\n=1}^{\infty}(\n L_{T})^{2}\cos(\n\theta) e^{-\p L_{T}^{2}}}{1+2\sum_{\n=1}^{\infty}\cos(\n\theta)e^{-\p L_{T}^{2}}}\,,
\end{align}
the mixing entropy per particle scales exponentially with system size similarly to the free energy (per particle).

\begin{figure}[t!]
\centering
\includegraphics[width=0.45\textwidth]{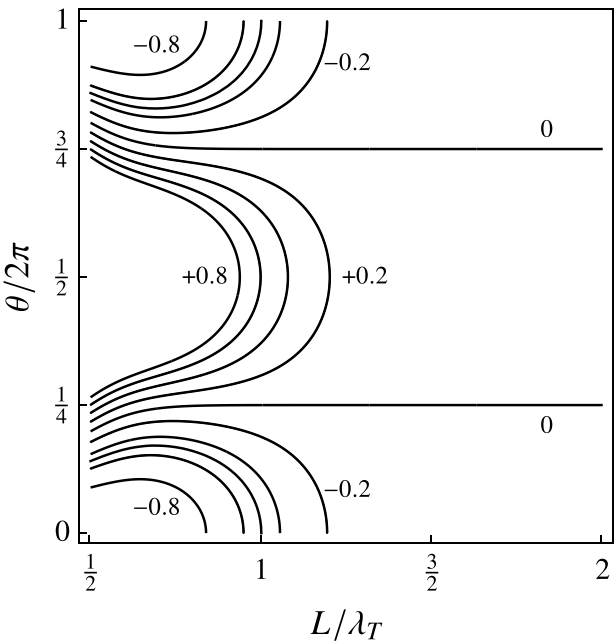}
\caption{
Contour plot of finite size correction to the average energy (normalized by its thermodynamic limit counterpart), as function of box size, $L/\l_{T}$ and the twist angle, $\theta$. (see text for discussion).}
\label{fig:energy}
\end{figure}

\subsection{Average Energy and Energy Fluctuations}
\label{sec:average-energy}

The (average) energy, unlike the free energy and entropy,
\begin{align}
\ol{E}(\theta)&=-\frac{\partial}{\partial \be}\log \c{Z}(\theta)\,, 
\end{align} 
is agnostic to the $N!$ in the partition function due to normalization by $\c{Z}(\theta)$.
Utilizing the single particle partition allows for a convenient extraction of not only the average energy in the thermodynamic limit
\begin{align}
\ol{E}(\theta)=\frac{d}{2}Nk_{B}T+\ol{E}_{L}(\theta)
\end{align}
but also finite volume corrections encoded in the latter term, $\ol{E}_{L}(\theta)$. The finite volume correction depends on the twist angle as
\begin{align}
\hspace{-3mm}
\frac{\ol{E}_{L}(\theta)}{\lim_{L\rightarrow\infty}\ol{E}(\theta)}=-\frac{4\p \sum_{\n=1}^{\infty}(\n L_{T})^{2}\cos(\n\theta)e^{-\p(\n L_{T})^{2}}}{1+2\sum_{\n=1}^{\infty}\cos(\n\theta)e^{-\p (\n L_{T})^{2}}}.
\end{align}
The denominator, modulo a factor of $L_{T}$, is the (one-dimensional) single particle partition function.
The finite volume average energy compared to its thermodynamic limit counterpart is
\begin{align}
\nonumber
&\frac{\ol{E}_{L}(\theta)}{\lim_{L\rightarrow\infty}\ol{E}(\theta)}\\
\nonumber
&=4\p\left[\,\left\{L_{T}^{2}\,e^{-2\p L_{T}^{2}}+\c{O}(L_{T}^{2}e^{-4\p L_{T}^{2}})\right\}\right.\\
\nonumber
&\left.+\cos\theta\left\{-L_{T}^{2}\,e^{-\p L_{T}^{2}}+\c{O}(L_{T}^{2}e^{-3\p L_{T}^{2}})\right\}\right.\\
\label{eq:EL-theta}
&\left.+\cos2\theta\left\{L_{T}^{2}\, e^{-2\p L_{T}^{2}}+\c{O}(L_{T}^{2}e^{-6\p L_{T}^{2}})\right\}+\cdots\right].
\end{align}
The expression has been organized to highlight finite size corrections for $\theta\neq\frac{\p}{2},\frac{3\p}{2}$ and $\theta=\frac{\p}{2},\frac{3\p}{2}$ -- in the latter cases, the leading contribution arises through the term proportional to $\cos2\theta$. The sign of the  correction depends on the twist angle as illustrated in Fig.~\ref{fig:energy}. There exist two critical values of the twist angle $\theta_{1}$ and $\theta_{2}$ at each box size for which the finite size correction is zero. For large box sizes $\theta_{1}$ asymptotes to $\frac{\p}{2}$ while $\theta_{2}$ asymptotes to $\frac{3\p}{2}$. For $\theta<\theta_{1}$ and $\theta>\theta_{2}$, the sign of the correction is negative. For $\theta_{1}<\theta<\theta_{2}$, however, the correction is positive.  

\begin{figure}[t!]
\centering
\includegraphics[width=0.45\textwidth, angle=0]{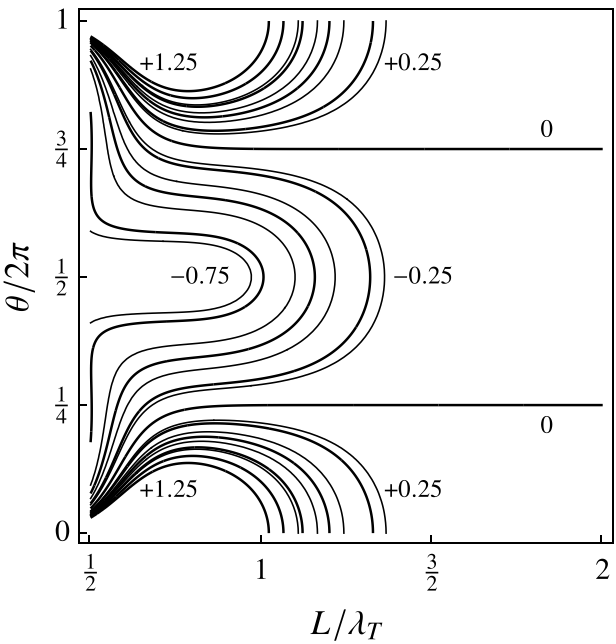}\hspace{40pt}
\caption{
Plot of the standard deviation normalized by the thermodynamic limit result as function of box size, $L/\l_{T}$ and the twist angle, $\theta$. (see text for discussion).}
\label{fig:stdev}
\end{figure}

Fluctuations around the average energy is characterized by the standard deviation, $\s_{E}$, 
\begin{align}
\s_{E}^{2}=\ol{E^{2}}-\ol{E}^{2}=\frac{\partial^{2}\log Z}{\partial\be^{2}}\,
\end{align}
For free particles, the width admits a Gaussian structure
\begin{align}
{{\s_{E}}/{\ol{E}}}&=\sqrt{\frac{2}{dN}}\,,
\end{align} 
as the `particles' in the gas do not interact and remain  entirely uncorrelated. The scaling with $N$ and $d$ are identical suggesting statistical independence between the $N$ particles in each of the $d$ dimensions. The boundary introduces statistical correlations due to particle-wall interactions~\cite{adhikari2025approaching}. The particular quantity of interest is the deviation from a perfectly uncorrelated gas, characterized by
\begin{align}
\d(\theta)&=\frac{\s_{E}(\theta)\,/\,\ol{E}(\theta)}{\sqrt{\frac{2}{Nd}}}-1\,.
\end{align}
A non-zero $\d(\theta)$ characterizes the amount of correlation in the gas present due to the boundary, which is encoded through $\ol{E}_{L}(\theta)$ -- performing an expansion in the limit of large $L_{T}$ using Eq.~\eqref{eq:EL-theta} leads to 
\begin{align}
\nonumber
\d(\theta)&=-\p^{4}L_{T}^{8}e^{-2\p L_{T}^{2}}+\c{O}(L_{T}^{6}e^{-2\p L_{T}^{2}})\\
\nonumber
&+\cos\theta\left[2\p^{2}L_{T}^{4}\,e^{-\p L_{T}^{2}}+\c{O}(L_{T}^{12}e^{-3\p L_{T}^{2}})\right]\\
&+\cos2\theta\left[-\p^{4}L_{T}^{8}\,e^{-2L_{T}^{2}}+\c{O}(L_{T}^{6}e^{-2\p L_{T}^{2}})\right]+\cdots\,.
\end{align}
The exact result for $\d(\theta)$ is plotted in Fig.~\ref{fig:stdev}.
For critical values of $\wt{\theta}_{1}$ and $\wt{\theta}_{2}$ there are zero finite size correlations. The width of statistical fluctuations around the average energy is larger than a Gaussian for $\theta<\wt{\theta}_{1}$ and $\theta>\wt{\theta}_{2}$ while the fluctuations are smaller for $\wt{\theta}_{1}<\theta<\wt{\theta}_{2}$. Finite size corrections in $\d$, compared to the average energy, are more acute compared to the average energy -- the features $\wt{\theta}_{1}<\theta_{1}$ and $\wt{\theta}_{2}>\theta_{2}$ are most readily evident for smaller box sizes.
\subsection{Pressure and Equation of State}
Finally, we consider finite size corrections to the pressure, defined as the infinitesimal change in the free energy due to an infinitesimal change in the system size,
\begin{align}
P_{d}(\theta)=-\frac{\partial F(\theta)}{\partial L^{d}}.
\end{align}
Rewriting the definition in terms of the partition function and a partial derivative in terms of $L$ is insightful
\begin{align}
P_{d}(\theta)=\frac{1}{dL^{d-1}}\frac{1}{\be}\frac{\partial}{\partial L}\log \c{Z}(\theta)\,.
\end{align}
The form suggests a connection to the (previously analyzed) average energy. Since $\c{Z}$ depends on $L$ and $\be$ through $\a$, 
\begin{align}
\a=\frac{\be\hbar^{2}}{2m}\left(\frac{2\p}{L}\right)^{2}
\end{align}
the pressure can be recast in terms of the energy (density)
\begin{align}
P_{d}(\theta)=-\frac{1}{d}\frac{L}{\be}\frac{\frac{\partial\a}{\partial L}}{\frac{\partial\a}{\partial\be}}\frac{\ol{E}(\theta)}{L^{d}}.
\end{align}
Utilizing the ratio of the partial derivatives, $\frac{\partial\be}{\partial L}=-\frac{2\be}{L}$,
leads to the finite volume ideal gas equation of state 
\begin{align}
\frac{P_{d}(\theta)}{\ol{E}(\theta)/L^{d}}=\frac{2}{d},
\end{align}
which is entirely independent of $\theta$. This is in spite of the fact that the pressure itself depends on $\theta$. This feature of the equation of state is not universal -- the commonly utilized Dirichlet boundary violates the identity.

\section{Conclusion}
\label{sec:conclusion}
In this manuscript, we have provided an accessible and systematic technique for studying the approach to the thermodynamic limit of a non-relativistic ideal gas (in a periodic box), which is a staple of undergraduate statistical mechanics. The key mathematical ingredient is the Fourier series, which is presented in Appendix~\ref{app:fourier-series} for convenience and readily accessible to an undergraduate audience. We have studied using the canonical ensemble the properties of an ideal gas. In particular, we have constructed finite size corrections to the free energy (per particle) and discussed its implications for the mixing entropy. Additionally, we investigated the average energy and its fluctuations in the energy near the thermodynamic limit. Remarkably, there are critical values of the twist angle for which finite size corrections are absent. We also investigated the connection of the pressure to the average energy and found that while both the energy and pressure depend on the twist angle and box size, the equation of state, their ratio, admits no finite size corrections. We leave the audience with practice exercises for review and mastery of the content.

\subsection{Example Exercises}
\label{app:example-sheet}
\begin{enumerate}
\item \textit{Dirichlet Partition Function.} The Dirichlet partition function has the form
\begin{align}
\wt{Z}_{1}&=\sum_{n=1}^{\infty} e^{-\be \e_{n}}&
\e_{n}&=\frac{\hbar^{2}k_{n}^{2}}{2m}\\
k_{n}&=\frac{\p n }{L}&
n&\in \mathbb{Z}^{+}\ .
\end{align}
\begin{enumerate}
\item Write $Z_{1}$ in terms of an infinite sum, $\sum_{n=-\infty}^{\infty} e^{-\be \e_{n}}$, by utilizing $\e_{n}=\e_{-n}$.
\item Show that the Dirichlet partition function admits the form 
\begin{align}
\wt{Z}_{1}=L_{T}-\tfrac{1}{2}+\D \wt{Z}_{1}\ .
\end{align}
\item Find the leading contribution to $\D \wt{Z}_{1}$.
\item Discuss the modification of $\wt{Z}_{1}$ in the presence of a zero ($n=0$) mode, i.e. $\e_{0}=0$.
\end{enumerate}

\item Using \texttt{Mathematica}, plot Fig~\ref{fig:Z} and add an analogous contribution to the ratio, $(L/\l_{T})/\wt{Z}_{1}$. Compare the convergence of the Dirichlet partition function to its periodic counterpart.
\item Find the contribution to the mixing entropy of two identical gases in a Dirichlet box. For each gas, assume $N$ particle and confinement into partitions, each with volume $L^{d}$.
\item For Dirichlet boundary condition, compare the finite size correction due to $N!$ in the free energy with that from the boundary. Identify the dominant correction.
\item Find the ratio
\begin{align}
\frac{P_{d}}{\ol{E}/L^{d}}
\end{align}
for an ideal gas in a Dirichlet box.
\item The Fourier transform of a periodic $\d$-function can be utilized to construct the \textit{Poisson summation formula}, 
\begin{align}
\sum_{n=-\infty}^{\infty}e^{-\a n^{2}}=\sqrt{\frac{\p}{\a}}\sum_{n=-\infty}^{\infty}e^{-\frac{\p^{2}}{\a}n^{2}}\,.
\end{align}
The formula allows for the restructure of the partition function that makes finite size effects in thermodynamic observables accessible.
\begin{enumerate}
\item Perform the integral  to exhibit the trivial identity 
\begin{align}
&\sum_{n=-\infty}^{\infty}e^{-\a n^{2}}\\
=&\sum_{n=-\infty}^{\infty}\int_{-\infty}^{\infty} d\k\, e^{-\a \k^{2}}\d(\k-n)\,.
\end{align}
Use the identity of Eq.~\eqref{eq:identity} to prove the summation formula.
\end{enumerate}
\end{enumerate}
\acknowledgments
P.A. is indebted to Brian Tiburzi for invaluable discussions and collaboration on related work.
P.A., and R.S acknowledge the support of the U.S. Department of Energy, Office of Science, Office of Nuclear Physics and Quantum Horizons Program under Award Number DE-SC0024385. The authors are indebted to Meifeng Lin for a careful reading of the manuscript. S.B. acknowledges the support of internal funds through Collaborative Undergraduate Research and Inquiry. P.A. acknowledges the hospitality of Brookhaven National Laboratory, and the support of the Kavli Institute for Theoretical Physics, Santa Barbara, through which the research was supported in part by the National Science Foundation under Grant No. NSF PHY-1748958.

\appendix
\section{Fourier Transform of a periodic function}
\label{app:fourier-series}
Consider a periodic function, $f(x)$, with period $L$,
\begin{align}
\label{eq:fx}
f(x)=\sum_{n=-\infty}^{+\infty}\d(x-nL)\,.
\end{align}
It can be recast as a Fourier sum
\begin{align}
f(x)=\sum_{\n=-\infty}^{+\infty}c_{\n}\,e^{\frac{2\p i \n x}{L}}\,.
\end{align}
The coefficient is determined through the orthogonality relation
\begin{align}
\int_{x_{0}}^{x_{0}+L}dx\,e^{\frac{2\p i \n x}{L}}e^{-\frac{2\p i \n' x}{L}}=L\, \d_{\n\n'}\,.
\end{align}
For $\n'=\n$, the integrand is one and the integral trivial. For $\n'\neq \n$, however, the integrand is a sinusoidal function that repeats at least once over the range of integration of size $L$. Consequently, the integral vanishes. The Fourier coefficient, $c_{\n}$, is determined by the orthogonality relation
\begin{align}
c_{\n}=\frac{1}{L}\int_{x_{0}}^{x_{0}+L}dx\,f(x)e^{-\frac{2\p i \n x}{L}}\,.
\end{align}
For the periodic function of interest, Eq~\eqref{eq:fx}, the coefficient is determined by the integral
\begin{align}
&\phantom{=}\frac{1}{L}\int_{x_{0}}^{x_{0}+L}dx\sum_{n=-\infty}^{\infty}\d(x-nL)\,e^{-\frac{2\p i \n x}{L}}\\
&=\frac{1}{L}\int_{x_{0}}^{x_{0}+L}dx\,\d(x-n_{0}L)\,e^{-\frac{2\p i \n x}{L}}\,.
\end{align}
In the second line, the sum over $\d$-functions has been replaced by a single delta function characterized by an integer $n_{0}$. The choice ensures that $n_{0}L$ lies within the range of integration. The resulting Fourier coefficients ares
\begin{align}
c_{\n}=\frac{1}{L}\,.
\end{align}
Therefore, the following identity has been established
\begin{align}
\sum_{n=-\infty}^{+\infty}\d(x-nL)=\frac{1}{L}\sum_{\n=-\infty}^{+\infty}e^{\frac{2\p i \n x}{L}}\,.
\end{align}
Performing a change of variables, $x\rightarrow xL$, and scaling out the period $L$ from the left-hand side leads to the dimensionless relation
\begin{align}
\label{eq:identity}
\sum_{n=-\infty}^{+\infty}\d(x-n)=\sum_{\n=-\infty}^{+\infty}e^{{2\p i \n x}}\,.
\end{align}
We utilize the identity in the construction of the exact partition function in Section~\ref{sec:reformulation-of-Z1}.

\bibliography{bib}


\end{document}